\documentclass[prb,preprint,superscriptaddress]{revtex4}
\usepackage{graphicx}
\usepackage{amssymb,amsmath}
\usepackage{bbold}
\newcommand{\plin}{$P_{lin}$}
\newcommand{\pix}{$\pi^X$}

\begin{document}
\title{Temperature insensitive optical alignment of the exciton in nanowire embedded GaN Quantum Dots} 
\author{A. Balocchi}
\email{andrea.balocchi@insa-toulouse.fr}
\affiliation{Universit\'{e} de Toulouse, INSA-CNRS-UPS, LPCNO, 135 avenue de Rangueil, 31077 Toulouse, France}
\author{J. Renard}
\affiliation{Department of Physics and Astronomy, University of British Columbia,
Vancouver, British Columbia, V6T 1Z4, Canada}
\affiliation{CEA-CNRS group "Nanophysique et semiconducteurs",
CEA-Grenoble, INAC-SP2M and CNRS-Institut N\'eel, rue des martyrs, 38054 Grenoble, France}
\author{C. T. Nguyen}
\affiliation{Universit\'{e} de Toulouse, INSA-CNRS-UPS, LPCNO, 135 avenue de Rangueil, 31077 Toulouse, France}
\author{B. Gayral}
\affiliation{CEA-CNRS group "Nanophysique et semiconducteurs",
CEA-Grenoble, INAC-SP2M and CNRS-Institut N\'eel, rue des martyrs, 38054 Grenoble, France}
\author{T. Amand}
\affiliation{Universit\'{e} de Toulouse, INSA-CNRS-UPS, LPCNO, 135 avenue de Rangueil, 31077 Toulouse, France}
\author{H. Mariette}
\affiliation{CEA-CNRS group "Nanophysique et semiconducteurs",
CEA-Grenoble, INAC-SP2M and CNRS-Institut N\'eel, rue des martyrs, 38054 Grenoble, France}
\author{B. Daudin}
\affiliation{CEA-CNRS group "Nanophysique et semiconducteurs",
CEA-Grenoble, INAC-SP2M and CNRS-Institut N\'eel, rue des martyrs, 38054 Grenoble, France}
\author{G. Tourbot}
\affiliation{CEA-CNRS group "Nanophysique et semiconducteurs",
CEA-Grenoble, INAC-SP2M and CNRS-Institut N\'eel, rue des martyrs, 38054 Grenoble, France}
\affiliation{CEA-LETI Minatec, 17 rue des martyrs, 38054 Grenoble, France}
\author{X. Marie}
\affiliation{Universit\'{e} de Toulouse, INSA-CNRS-UPS, LPCNO, 135 avenue de Rangueil, 31077 Toulouse, France}
\date{\today}
\begin{abstract}
We report on the exciton spin dynamics of nanowire embedded GaN/AlN Quantum Dots (QDs) investigated by time-resolved photoluminescence spectroscopy.
Under a linearly polarized quasi-resonant excitation we evidence the quenching of the exciton spin relaxation and a temperature insensitive degree of the exciton  linear polarization, demonstrating the robustness of the optical alignment of the exciton spin in these nanowire embedded QDs. A detailed examination of the luminescence polarization angular dependence shows orthogonal linear exciton eigenstates with no preferential crystallographic orientation.   
\end{abstract}
\maketitle
\begin{figure}
 \centering
 \includegraphics[width=0.5\textwidth,keepaspectratio=true]{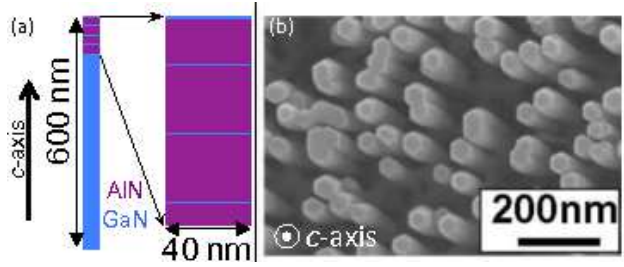}
 \caption{(Color online) (a) Schematic representation of a single nanowire. (b) Top view SEM image of the nanowires.} 
 \label{figure_SEM}
\end{figure}
\begin{figure}
 \centering
 \includegraphics[width=0.5\textwidth,keepaspectratio=true]{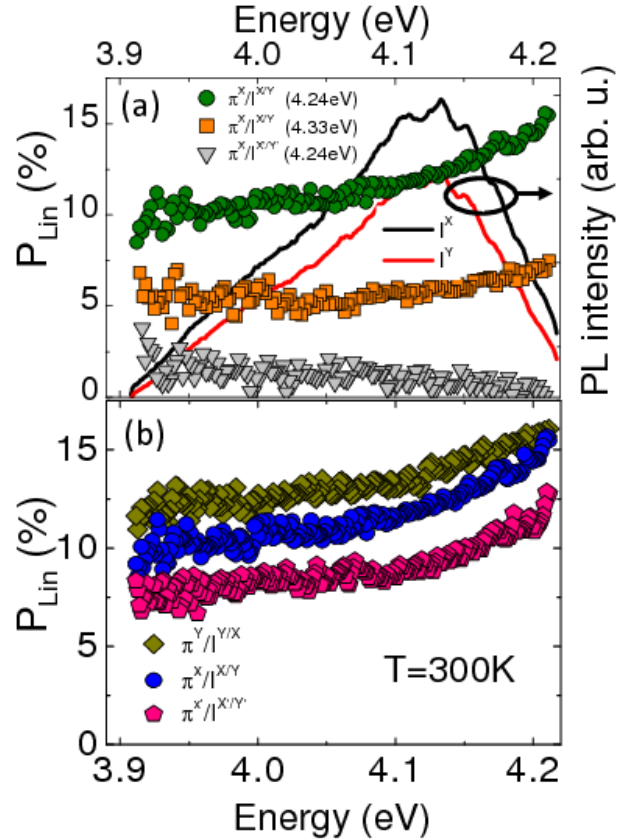}
 \caption{(color online) (a) The co- ($I^X$) and cross-polarized ($I^Y$) time integrated PL intensities (solid lines) following a  quasi-resonant excitation ($E_{exc}$=4.24 eV) at T=300 K. The PL linear polarization degrees following different configuration of linear excitation and detection are also plotted. $\pi^{X(Y)}$  indicate respectively a linearly (X or Y, with respect to the laboratory frame)  polarized excitation. The superscripts in $I^{X/Y}$ ($I^{X'/Y'}$) indicate the
 polarization directions  of the PL intensities  used to calculate the PL linear polarization degree. (b) The  PL linear polarization degree measured after a simultaneous rotation of the excitation and detection reference frames (E$_{exc}$= 4.24 eV).}
 \label{figure_2}
\end{figure}
\begin{figure}
\centering
 \includegraphics[width=0.5\textwidth,keepaspectratio=true]{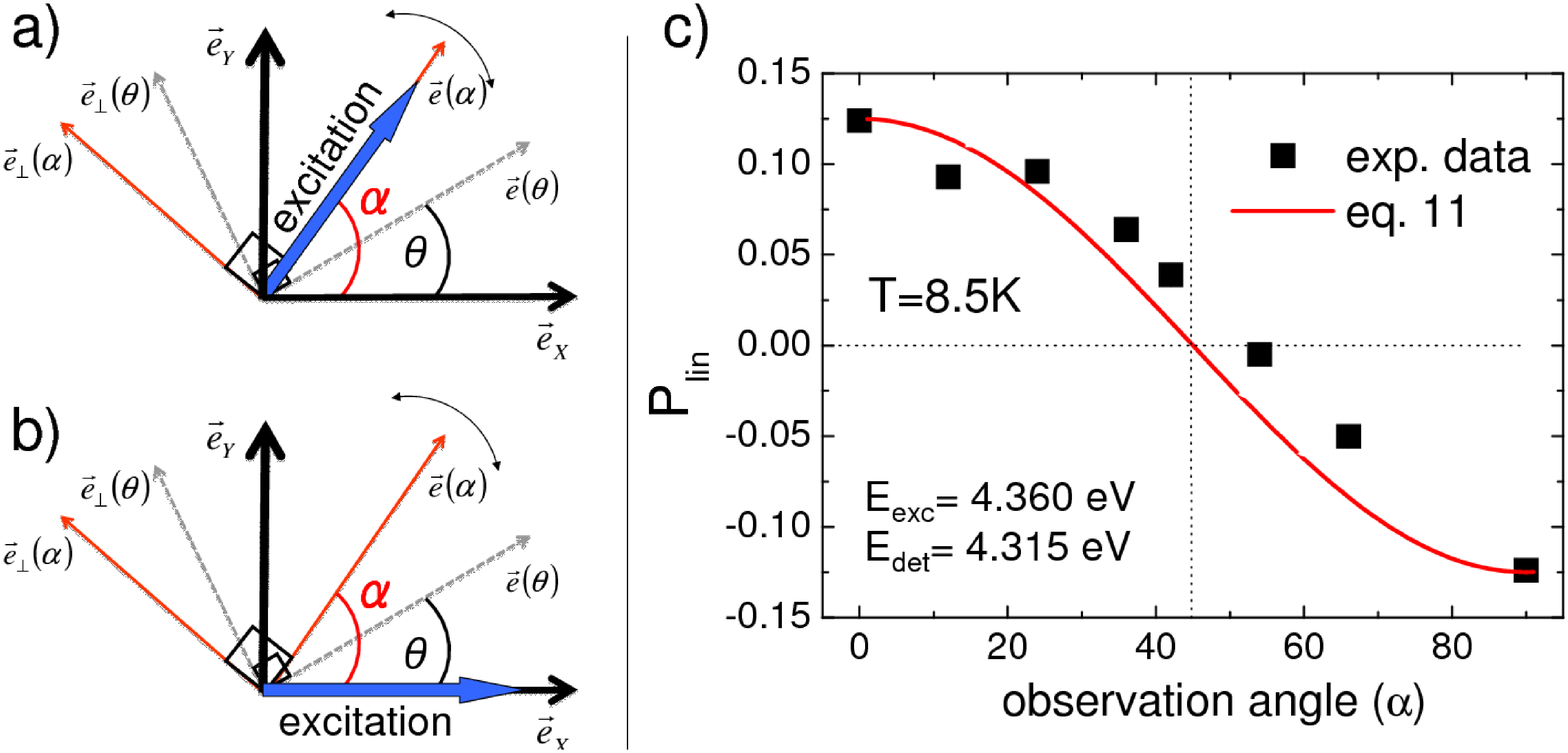}
 \caption{color online. a), b) : Geometry of the experiment A and B respectively. c) The Linear polarization degree of the photoluminescence following a linear excitation along the $\vec{e}_X$ axis, measured along the $\left(\vec{e}(\alpha), \vec{e}_{\perp}(\alpha)\right)$ basis (experiment B, T=8.5 K). The red line is a fit to the data according to equation~\ref{eq:depolarization_angle} (see text). }
 \label{fig:depolarization}
\end{figure}
\begin{figure}
 \centering
 \includegraphics[width=0.5\textwidth,keepaspectratio=true]{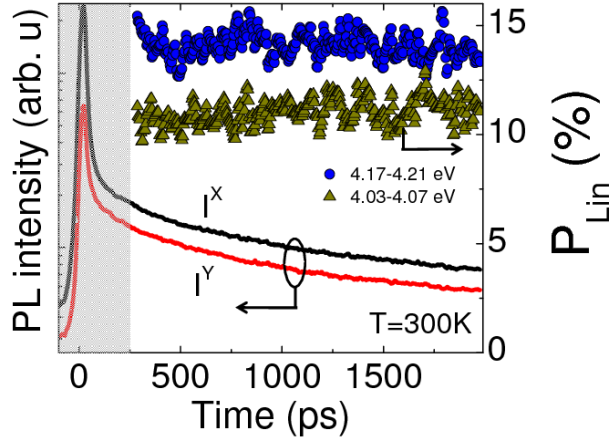}
 \caption{Time evolution of the PL components co- ($I^X$) and counter-polarized ($I^Y$) with the linearly polarized quasi resonant excitation (E$_{exc}$=4.24 eV)
 and the corresponding linear polarization degree measured at two different energies. The gray area represents the temporal region dominated by the laser scattered light.}
 \label{fig:dynamics}
\end{figure}
\begin{figure}
 \centering
 \includegraphics[width=0.5\textwidth,keepaspectratio=true]{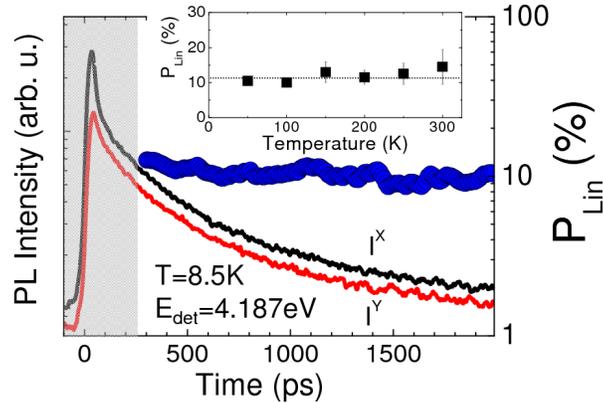}
 \caption{ Dynamics of the low temperature (T=8.5 K) PL components co- ($I^X$) and cross-polarized ($I^Y$) with the linearly polarized  excitation and the corresponding linear polarization degree. The gray area represents the temporal region dominated by the laser scattered light. Inset: the measured PL linear polarization degree for all the temperatures investigated. The energy difference between excitation and detection
 has been kept constant at $\Delta E=$133 meV, and the excitation energy has been varied according to the temperature variation of the
QD gap  in order to excite the same nanowire family.} 
 \label{fig:temperature}
\end{figure}

\label{intro}
Wide gap GaN and  its alloys have proven to be  a very successful and versatile system for a variety of optoelectronics applications ranging from light emission 
 (lasers or light-emitting diodes) ~\cite{Nakamura1994,Nakamura1996,feltin_broadband_2009,christmann_room_2008} to photovoltaic conversion devices~\cite{Wu2003}.
More recently, bulk or nanostructured GaN has emerged as well as a suitable material system for controlling and manipulating the exciton spin up to room temperature~\cite{beschoten_spin_2001,tackeuchi_nanosecond_2006,bu_long_2010,bu_temperature_2010,nagahara_no_2006} thanks to  its  very low spin-orbit coupling and wide band-gap~\cite{Dyakonov_book,vurgaftman_band_2003,fu_spin-orbit_2008}. The tri-dimensional confinement obtained in nanostructures like quantum dots or nanoparticles is expected to further improve the stability of the carriers' spin memory by reducing the efficacy of the wave vector dependent spin scattering mechanisms~\cite{Dyakonov_book}. Despite these strong potentialities, very few experimental investigations have been performed on the exciton spin dynamics of (In)GaN Quantum Dots (QDs). Lagarde $et~al.$~\cite{Lagarde2008} have measured the optical alignement
of the exciton in zinc-blende GaN/AlN dots up to room temperature, while S\'en\`es $et~al.$~\cite{Senes2009} have demonstrated the control of the exciton linear polarization degree through the application of an external electrical bias in a 	wurtzite InGaN/GaN  p-i-n QD structure. High density of defects and dislocations however still affects the growth of (Al)(In)GaN materials, due to the lack of proper substrates.  This condition hampers both the optical and spin properties of these compounds~\cite{Brimont2008}. Nanowire-embedded GaN/AlN QDs present a valid alternative to Stranski-Krastanov (SK)  grown nanostructures thanks to their higher  crystalline quality due to the reduced interaction with the substrate and the consequent lower density of defects near the dot site~\cite{Songmuang2007}. In addition, these wurtzite nanowires grow with the crystallographic $c$ axis oriented normal to the silicon  (111) substrate surface, offering a new interesting system to study the polarization properties of confined excitons in high quality nitrides QDs by optical orientation experiments~\cite{Renard2009,Renard2008}.\\
 In this work, we report on a detailed time-resolved investigation of the exciton spin dynamics of an ensemble of wurtzite GaN/AlN QDs embedded in nanowires  up to room temperature. By means of optical orientation experiments  we demonstrate that the exciton photoluminescence (PL) linear polarization degree arises from the optical alignment of the exciton spin  exhibiting neither a temporal nor an amplitude decay up to room temperature. The temperature insensitivity of the  optical alignment  of the excitons contrasts with what was previously reported for zinc-blende GaN/AlN  and wurtzite InGaN/GaN  SK dots~\cite{Lagarde2008,Senes2009}. Finally, the angular analysis of the PL linear polarization evidences exciton orthogonal linear eigenstates with no preferential in-plane orientation.\\
 \\
 \label{sample&exp}
The sample under study consists of an ensemble of GaN/AlN Quantum Dots embedded in $\approx$ 40 nm diameter nanowires containing 10 planes of $\approx$ 1 nm  height GaN QDs with  a nanowire density of $\approx$ 10$^{10}$cm􀀀$^{-2}$ (figure~\ref{figure_SEM}). The nanowires are about 600 nm long, the AlN/GaN heterostructure being grown on top of a 500 nm long base of GaN. The nanowires were grown on a Si (111) substrate and the details of growth conditions can be found in reference~\onlinecite{Songmuang2007}. The excitation source is provided by a mode-locked frequency-tripled Ti:sapphire laser, with a 1.5 ps pulse width and a tunable energy  in the range 4.00-4.77 eV. The laser beam, propagating parallel to  the nanowire $c$ axis, is focused onto the sample to a 50 $\mu$m diameter spot with a time averaged power in the range P$_{exc}$=0.1-1 mW. The PL signal is then dispersed by an imaging spectrometer and it is temporally resolved by a S20 photocathode streak camera with an experimental set-up time resolution of 8 ps. The linear polarization degree of the luminescence is defined as $P_{lin}=(I^{\alpha}-I^{\alpha_{\perp}}) / (I^{\alpha}+I^{\alpha_{\perp}})$. Here $I^{\alpha}$ and $I^{\alpha_{\perp}}$ denote 
the linearly polarized PL intensity components with polarization parallel to the Cartesian orthogonal frame axis defined by the $\left(\vec{e}(\alpha), \vec{e}_{\perp}(\alpha)\right)$ basis rotated by and angle $\alpha$ with respect to the laboratory frame $\left(\vec{e}_X, \vec{e}_Y\right)$ and chosen so that the $X-Y$ plane is perpendicular to the sample growth direction. For $\alpha$=0, $\vec{e}(\alpha)\equiv \vec{e}_X$ and $\vec{e}_{\perp}(\alpha)\equiv \vec{e}_Y$.
\\
\label{results_wavelength}
Figure~\ref{figure_2}(a) (solid lines) reports the time-integrated components of the PL intensity co- ($I^{X}$) and cross-polarized ($I^{Y}$) with the quasi-resonant linear excitation set along $\vec{e}_X$ (\pix, $E_{exc}$=4.24 eV) at T=300 K with the corresponding linear polarization degree (circles). Quasi-resonant excitation means here that the laser excitation
energy is set within the energy range spanned by the ensemble of QD PL. A distinct PL linear polarization degree is observed for any detection energy in the emission spectrum, with larger values \plin$\approx$ 15 \% in the high-energy part. This trend is confirmed by the overall marked decrease of the PL linear polarization degree after a \pix~laser excitation of higher energy (figure~\ref{figure_2}(a), squares). For a linearly polarized laser excitation $E_{exc}\gtrsim$ 450meV above the detection, no PL linear polarization is measured within the experimental uncertainties (not shown).
The PL  linear polarization degree measured along two orthogonal axis (X$'$,Y$'$) rotated by $\alpha$=$\pi$/4 with respect to a quasi resonant linear excitation parallel to the laboratory reference frame (\pix)  is also reported in figure~\ref{figure_2}(a) (downwards triangles). No linear polarization degree is observed (\plin$^{\alpha=\pi/4}\leq$2 \%  on the whole PL spectrum) suggesting the existence of linear orthogonal excitonic states.
\\We have completed the previous results with a careful investigation of the angular dependence of the exciton PL polarization.
We have first measured the PL linear polarization along two orthogonal axis $\left(\vec{e}(\alpha), \vec{e}_{\perp}(\alpha)\right)$  as a function of a simultaneous in plane rotation of the excitation polarization  and detection linear polarization frame (experiment A, figure~\ref{fig:depolarization}-a) with $\pi^{\alpha} \parallel \vec{e}_{\alpha}$ . Photoluminescence co-polarized with the linear exciting laser is always measured. Figure~\ref{figure_2}(b) shows, as an example, the measured PL linear polarization degree after an excitation along the $X$ axis (circles), the $Y$ axis (diamonds) and along the X$'$ axis (pentagons). We have obtained quantitatively very similar results for any temperatures and for an arbitrary rotation $\alpha$ of the excitation and detection frames around the crystallographic $c$ axis indicating the existence of different families of nanowire-embedded QDs characterized by linear orthogonal excitonic eigenstates with quasi-random in-plane orientation. The small difference observed in the linear polarization degree ($\Delta P_{lin}\approx$3\%), might be attributed to slightly preferred orientations of the orthogonal linear exciton eigenstates in the sample plane.\\
In the second case, experiment B (figure~\ref{fig:depolarization}-b), we have kept fixed the laser linear excitation along $\vec{e}_X$ and we have measured the 
PL linear polarization degree in the $\left(\vec{e}(\alpha), \vec{e}_{\perp}(\alpha)\right)$ basis. We have restricted $\alpha$ to the range $0\leq \alpha < \pi/2$ since \plin ($\alpha +\pi/2)=$ -\plin ($\alpha$). Figure~\ref{fig:depolarization}-c, reports the experimental data of the time integrated \plin~ for an energy separation between the excitation and
 detection $\Delta E_{e-d}$=45 meV~\cite{note2}. These results as a function of the observation angle confirm the optical creation of linearly polarized exciton states with random orientation.\\ 
The ensemble of the  observations reported in figures~\ref{figure_2} and \ref{fig:depolarization}, together with the absence of PL circular polarization under circular excitation (not shown), are consistent with the optical alignment of linearly polarized excitons in these wurtzite GaN/AlN nanowires. The optical  alignment of the exciton has been already reported for many III-V or II-VI bulk materials or heterostructures~\cite{Paillard2001,Tsitsishvili2002,Bonnot1974,Meier1984,Amand1997}. In QDs, the origin of the linear symmetry of the exciton eigenstates  is generally attributed to the anisotropic electron-hole exchange splitting~\cite{Gammon1996, bayer_2002,kindel_exciton_2010} originating from QD elongation and/or interface anisotropy. Here, the symmetry of the dots is generally $C_{3v}$~\cite{tronc_symmetry_2004} , so the interface anisotropy argument is not relevant. Previous results for III-N SK quantum dots reported definite directions for the orientation of the splitted exciton orthogonal eigenstates. These directions were linked to the symmetry of the semiconductor crystal for zinc-blende structures~\cite{Lagarde2008} (orientation along the [110]/[1$\bar 1$0] directions) and to a symmetry reduction probably due to a residual substrate anisotropy for wurtzite structures~\cite{Senes2009} (orientation along a subset of [1$\bar 1$00]/[11$\bar 2$0] directions). X-ray analysis performed on similarly grown nanowire embedded QDs has however revealed the same crystallographic orientation for the nanowires corresponding to an in-plane coincidence of the GaN and Si lattices~\cite{Largeau2008}. Therefore, either an in-plane asymmetry due to imperfect hexagonal facets or the roughness fluctuations at the GaN/AlN interfaces~\cite{Gammon1996} are likely at the origin of the observed random orientation of the linearly-polarized exciton eigenstates, whereas the arbitrary in-plane crystallographic orientation of the nanowires is ruled out~\cite{ivchenko_fine_1997}.\\
\label{results_dynamics}
Figure~\ref{fig:dynamics} presents the time evolution at room temperature of the co- and cross-polarized PL intensity components obtained after a quasi resonant  linearly polarized  excitation ($E_{exc}$= 4.24 eV, \pix) and the corresponding dynamics of the PL linear polarization degree measured at two different detection energies. Note that at short times, the PL is dominated by the fast decay of scattered laser photons (grayed out on the decay curve).  The short PL decay time ($\tau_{PL}\approx$ 1 ns) is  an indication of the weak contribution of the Quantum Confined Stark Effect in this structure thanks to the QD small height. The identical $\tau_{PL}$ observed for the $I^X$ and $I^Y$ components is  the signature of the comparable oscillator strength of each orthogonal exciton state. 
The QD emission dynamics exhibits a  polarization degree which remains strictly constant in time,  within our experimental accuracy, during the exciton lifetime for every  wavelength within the PL spectrum. Under a circularly polarized excitation, quantum beats at the pulsation corresponding to the anisotropic exchange splitting should be observed in the PL polarization dynamics. The absence of any measurable polarization (not shown) is attributed here to the exchange splitting energy statistical fluctuations among the QD ensemble. As we do not observe any beats for t $>$ 200 ps, we can infer that the standard deviation of the anisotropic exchange interaction distribution is larger than 10 $\mu$eV, which is consistent with recent measurements on SK QDs~\cite{kindel_exciton_2010} .\\
We have  studied the PL polarization dynamics as a function of the temperature. As the degree of the PL linear polarization depends on the excitation-detection energy separation $\Delta E_{e-d}$, to ensure the analysis of the same QD family, we have kept $\Delta E_{e-d}$ constant  while varying the excitation energy following a simple Varshni law for the band gap temperature dependence. Figure~\ref{fig:temperature} presents the low temperature (T=8.5 K) PL components co- ($I^X$) and cross-polarized ($I^Y$) with the linear excitation  and the corresponding linear polarization degree while the inset reproduces the value of the  linear polarization degree for all the measured temperatures up to room temperature.  PL linear polarization features very similar to those observed at 300 K are recorded for any temperature investigated. The very long spin relaxation time measured at low temperature are in agreement with recent calculations of the polarized exciton relaxation rates in GaN QDs~\cite{tong_theory_2011} with in-plane aspect-ratio close to 1. We can conclude that no temporal decay nor a decrease of the \plin~degree are observed for any temperature within our experimental uncertainties. 

Room temperature exciton optical alignement was recently  evidenced in cubic GaN/AlN~\cite{Lagarde2008} and wurtzite InGaN/GaN~\cite{Senes2009} self assembled QDs, and in  QD-like structures induced by In segregation in InGaN/GaN Quantum Wells~\cite{Nagahara2006}.
 In ref.~\onlinecite{Lagarde2008,Senes2009} despite a likewise time constant PL linear polarization degree  observed up to room temperature, a clear decrease ($\approx$ 50 \%) of the amplitude of the linear polarization was evidenced from T=10 K to T=300 K. An activation energy for the polarization degree was determined, for instance for zinc-blende GaN/AlN SK QDs~\cite{Lagarde2008}, in the range 50-100 meV.
The temperature insensitivity of the exciton alignment in figure~\ref{fig:temperature} is therefore a distinctive property of these nanowire embedded GaN/AlN QDs which strikingly singles out the present findings from the preceding results.
The origin of the temperature robustness of the  exciton alignment is still under investigation. Confinement effect comparable to SK GaN/AlN quantum dots are expected for this nanowire structure. This remarkable behavior has therefore to be linked to the peculiar structure of the nanowires. We tentatively ascribe the origin of the observed temperature stability to the higher structural quality of the nanowire embedded QDs compared to the SK structures. Different exciton-phonon interaction compared to SK dots could also play a non-negligible role thanks to the marked different structural configuration of the nanowires~\cite{tong_theory_2011,yin_electron_2010} .

 \label{sec:depol_discussion}
We have developed a simple model to describe the  measured PL linear polarization degree  according to the  experimentally deduced orientation and symmetry of the exciton dipoles in the nanowire embedded QDs. We first evaluate the maximum PL linear polarization degree measurable along the $\left(\vec{e}(\alpha), \vec{e}_{\perp}(\alpha)\right)$ basis,  following a quasi resonant excitation along  $\vec{e}_{\alpha}$ (experiment A). We postulate a random orientation of exciton dipoles corresponding to the orthogonal eigenstates identified by the $\left(\vec{e}(\theta), \vec{e}_{\perp}(\theta)\right)$ basis for a given QD, with $0\leq \theta <\pi$.\\ 
We first assume  that during the energy relaxation the relative phase between the exciton states is $fully$ preserved.
Assuming in addition that the oscillator strength of the two orthogonal exciton dipoles are roughly identical, the projection of the wavefunction on the bright excitonic states is:
\begin{equation}
\mid\psi_{\theta,\omega_1}(t)\rangle  _{\theta}=\left( e^{i\omega_1t/2} \cos \left( \theta - \alpha \right) \right) \mid X \rangle _{\theta} -\left (e^{-i\omega_1 t/2} \sin \left( \theta - \alpha\right) \right) \mid Y \rangle _{\theta}
\end{equation}
where $\hbar \omega_1$ is the fine structure splitting of the exciton bright states due to anisotropic exchange interaction (a global phase factor has been dropped for simplicity). In the basis $\left(\vec{e}(\alpha), \vec{e}_{\perp}(\alpha)\right)$ associated with the laser polarization we obtain:
\begin{equation}
\begin{split}
\mid\psi_{\theta,\omega_1}(t)\rangle  _{\alpha}=\left[ e^{i\omega_1t/2} \cos^2 \left( \theta - \alpha \right) +  e^{-i\omega_1 t/2} \sin^2 \left( \theta - \alpha\right) \right] \mid X \rangle _{\alpha} +\\
\left[ e^{i\omega_1t/2} \sin\left( \theta - \alpha \right) \cos \left( \theta - \alpha \right) -  e^{-i\omega_1 t/2}\sin\left( \theta - \alpha \right) \cos \left( \theta - \alpha \right)  \right] \mid Y \rangle _{\alpha}.
\end{split}
\end{equation}
The density matrix of the Quantum Dot ensemble is then given by:
\begin{equation}
\rho(t)=\int_0^{\pi}d\theta\int d\omega_1~g(\theta,\omega_1) ~\mid \psi_{\theta,\omega_1}(t)\rangle  _{\alpha}~\langle \psi_{\theta,\omega_1}(t)\mid _{\alpha}.
\end{equation}
We make here the further assumption that there is no correlation between the fine structure splitting amplitude and the orientation $\theta$ of the excitonic dipoles, so that we have: $g(\theta,\omega_1)=g_0(\theta)g_1(\omega_1)$.  We model then the angular distribution function $g_0(\theta)$ by a uniform law and the fine structure splitting distribution $g_1(\omega_1)$ by a Gaussian centered on $\bar{\omega}$ and with standard deviation $\sigma_{\omega}$, so that:
\begin{eqnarray}
g_0(\theta) & =  & 1/\pi \\
g_1(\omega_1) & =  & \frac{1}{\sqrt{2\pi} \sigma_{\omega}}e^{-\frac{(\omega_1-\bar{\omega})^2}{2 \sigma_{\omega}^2}}.
\end{eqnarray}
\\
In the investigated time range (t$>$200 ps), the condition t$\gg\sigma_{\omega}^{-1}$ holds and the density matrix writes at t$_0\approx$ 200 ps:
\begin{equation}
\rho(t_0)=\left( \begin{array}{cc}
3/4 & 0 \\
0 & 1/4
\end{array}
\right).
\end{equation}
As we do not observe any linear polarization relaxation during the exciton lifetime, the subsequent time evolution is given by
\begin{equation}
\rho(t)=\rho(t_0) e ^{-t/\tau_{rad}}
\end{equation}
where we have added the exponential factor to take into account the exciton radiative recombination. The PL linear polarization in the $\left(\vec{e}(\alpha),\vec{e}_{\perp}(\alpha)\right)$ basis is therefore $P_{lin}(t)=1/2$ which is constant and independent of the laser linear polarization direction.
Experimentally we obtain $P_{lin}^{X/Y}\leq$0.15 on the whole detection spectrum signifying that depolarization mechanisms occur during energy relaxation.\\In a more realistic approach we consider now that, during energy relaxation, the relative phase between exciton states is only $ partially$ preserved. We characterize  the coherence loss during the energy relaxation process by a phenomenological energy dependent coefficient  $C\left(\Delta E_{e-d}\right)$ so that $\rho(t_0)$ is replaced by:
\begin{equation}
\rho(t_0)^{eff} = \rho(t_0) \left(1-C\right)+ \left(C/2\right)\hat{\mathbb{1} }
\end{equation}
where $\hat{\mathbb{1}}$ is the identity matrix.
Experimentally we observe  in figure~\ref{fig:depolarization}-c  $P_{lin}\left(\alpha=0\right)$=12.5\% for a PL detection 45 meV below the excitation, implying $C\left(\mathrm{45~meV}\right)$=0.75. 
Note that a similar depolarization value was measured in InAs/GaAs quantum dots by comparing the PL polarization degree  obtained in strictly resonant excitation and quasi-resonant excitation (one LO phonon above the ground exciton state) conditions~\cite{senes_exciton_2005}.\\
We now turn to the modeling of experiment B, where the excitation linear polarization is kept fixed along the $\vec{e}_X$ direction while we rotate the observation basis 
$\left(\vec{e}(\alpha),\vec{e}_{\perp}(\alpha)\right)$ by an arbitrary angle $\alpha$.
The projection of the excitonic state on the bright state basis, just after the energy relaxation, is now given in the basis $\left\{ \mid X \rangle _{\alpha}, \mid Y \rangle _{\alpha} \right\}$ by:
\begin{equation}
\begin{split}
\mid\psi_{\theta,\omega_1}(t)\rangle  _{\alpha}=\left[ e^{i\omega_1t/2} \cos(\theta) \cos \left( \theta - \alpha \right) +  e^{-i\omega_1 t/2} \sin(\theta) \sin \left( \theta - \alpha\right) \right] \mid X \rangle _{\alpha} +\\
\left[ e^{i\omega_1t/2} \cos(\theta) \sin\left( \theta - \alpha \right) -  e^{-i\omega_1 t/2}\sin(\theta) \cos \left( \theta - \alpha \right)  \right] \mid Y \rangle _{\alpha}.
\end{split}
\end{equation}
Under the same assumption as previously detailed, we obtain, for t$>$200 ps:
\begin{equation}
\rho(t_0)=\frac{1-C}{2}\left( \begin{array}{cc}
1+\frac{\cos (2\alpha)}{2} & 0 \\
0 & 1-\frac{\cos (2\alpha)}{2}
\end{array}
\right)+ \frac{C}{2}\hat{\mathbb{1}}.
\end{equation}
The  PL linear polarization is now given by:
\begin{equation}
P_{lin}(\alpha,C)=\frac{\cos(2 \alpha)}{2}(1-C).
\label{eq:depolarization_angle}
\end{equation}
Figure \ref{fig:depolarization}-c presents the experimental data and the calculated curve according to  equation (\ref{eq:depolarization_angle}). The agreement is satisfactory considering that no oscillator strength difference between the exciton eigenstates nor any anisotropy of the exciton states distribution in the sample plane have been considered~\cite{nota}.\\
In conclusion, we have studied the optical orientation of the exciton spin in nanowire-embedded GaN/AlN Quantum Dots. We have observed  the quenching of the exciton spin relaxation time and  the temperature insensitivity of the exciton PL linear polarization degree. These results contrast  with the temperature  decrease of the linear polarization degree observed in other nitride based QD systems~\cite{Lagarde2008,Senes2009}. A careful investigation of the angular dependence of the PL linear polarization reveals the presence of independent families of QDs with orthogonal exciton eigenstates without any preferential crystallographic in-plane orientation.\\ Acknowledgments:  we gratefully acknowledge fruitful discussions with M. Wu and E. L. Ivchenko. We thank P.M. Chassaing for his contribution to the experiments. J.R acknowledges funding from the CIFAR JF Academy.


\begin{thebibliography}{36}
\bibitem[{not({\natexlab{a}})}]{note2}
\bibinfo{note}{These measurements were done at low temperature (T=8.5 K) in
  contrast with the results in figure 2 (T=300 K). As demonstrated later in the
  paper, the polarization properties in these GaN QDs are very weakly
  temperature sensitive.}
\bibitem[{not({\natexlab{b}})}]{nota}
\bibinfo{note}{We have checked that any distribution $g(\theta)$ of the exciton
  dipole orientation possessing the $C_{6v}$ symmetry does not modify the
  results}.
\end{thebibliography}
\end{document}